\documentclass[twocolumn,english,aps,pra,superscriptaddress,showpacs,tightenlines]{revtex4-1}
\usepackage{amsmath}
\usepackage{graphicx}
\usepackage{amssymb}
\usepackage{CJK}
\usepackage{color}

\begin{document}


\title{Controllable single-photon frequency converter via a one-dimensional
waveguide}

\author{Z. H. \surname{Wang} }
\affiliation{Beijing Computational Science Research Center, Beijing 100084,
China}
\author{Lan \surname{Zhou} }
\email{zzhoulan@gmail.com}
\affiliation{Key Laboratory of Low-Dimensional Quantum Structures and Quantum
Control of Ministry of Education, and Department of Physics, Hunan
Normal University, Changsha 410081, China}
\affiliation{Beijing Computational Science Research Center, Beijing 100084, China}
\author{Yong \surname{Li} }
\email{liyong@csrc.ac.cn}
\affiliation{Beijing Computational Science Research Center, Beijing 100084, China}
\author{C. P. \surname{Sun}}
\affiliation{Beijing Computational Science Research Center, Beijing 100084, China}

\begin{abstract}
We propose a single-photon frequency converter via a one-dimensional waveguide coupled
to a $V$-type atom. The on-demand classical field allows the atom to absorb a photon
with a given frequency, then emit a photon with a carried frequency different from
the absorbed one. The absorption and re-emission process is formulated as a two-channel
scattering process. We study the single-photon frequency conversion mechanism in two kinds of
realistic physical system:  coupled resonator waveguide with cosine
dispersion relation and an optical waveguide with linear dispersion relation respectively.
We find that the driving field prefers weak in coupled resonator waveguide but arbitrarily strong in
optical waveguide to achieve an optical transfer efficiency.
\end{abstract}

\pacs{03.65.Nk, 03.67.Lx, 78.67.An}
\maketitle


\section{Introduction}

Photon frequency conversion~\cite{Kumar,jmhuang,tucker} refers to
transducing the input photons with a given frequency into the output photons with
a different frequency while preserving the non-classical quantum properties. 
Experimentally, the photon frequency conversion has been achieved in the nonlinear
medium by frequency mixing technologies~\cite{Babdo,Zaske,MTR,SR,MW}.
In particular, Especially, the single-photon frequency conversion~\cite{Ates} has
many applications in quantum information and quantum communication process,
and the effective single-photon frequency conversion scheme~\cite{PS} has been
proposed in a waveguide channel with the assistance of the Sagnac interferometer which couples
to a multi-level emitter~\cite{WB,shen1,shen2}.

Recently, the coherent control of the single photon has been studied in the one-dimensional
(1D) waveguide with linear~\cite{Fan051} and nonlinear~\cite{Lan08,Plongo,ZH}
dispersion relations, where the two-level or three-level system acts
as a quantum switch. When the three-level system is applied as a scatter~\cite{Lan13,TT,ZYZhang}, the
photons can be transferred from one channel to another, and the
carried frequency in different channels can be same or different, depending
on whether the atom experiences the internal state transition in the
scattering process or not.
In present study, we aim to study a single-photon frequency converter mechanism 
in a single channel which is realized with waveguide.

To this end, we propose a scheme via a single waveguide with a located $V$-type atom, where the photon in the waveguide couples to one arm transition, and is
scattered by the atom. To convert the frequency of the incident photon, we
drive the other arm transition of the atom via a classical field.
The subsystem consisting of the classical driving and the related two levels coupling to it, forms a pair of non-generated dressed ground states, which supports two channels (but still in the same waveguide) for the incident photons. The energy
conservation implies that the frequency of the incident photons will be converted when it is
transferred from one channel to the other. We would remark that the $V$-type three-level system has been also used in
Ref.~\cite{Lan13}, where the carried frequencies of photons in two channels are the same because
only a single atom ground state is involved. Compared to the previous
schemes~\cite{Lan13}, our scheme has the following two advantages: (i) We would apply only one
single waveguide, which can be realized more easily with the current
experimental technologies. (ii) The frequency difference between the two
channels can be controlled only by adjusting the frequency and strength of the driving field,
instead of the atomic energy level configuration.

In this paper, we firstly give a general description for the single-photon
transmission and frequency conversion mechanism in one dimensional waveguide based on
Lippman-Schwinger equation~\cite{Taylor}. Then we consider two explicit
models: the coupled resonator waveguide (CRW)~\cite{Lan08,Lan13,Plongo,TT,ZH} and one-dimensional optical waveguide~\cite
{TST1,Fan07,DW}. For the case of CRW, the single-excitation spectrum of each of the two channels has the structure of one energy band and two discrete bound states with one of them being above the energy band and the other below it. On the contrary, the related spectrum structures of the two channels in the case of optical waveguide have only the lower energy bounds but no upper ones and bound states. In both of these models, the frequency converter only works when the incident photon frequency is inside the overlap regime of the spectra of the two channels. We
find that the classical driving field prefers to be weak in the CRW but arbitrarily strong in
optical waveguide to achieve an optimal transfer efficiency.

The rest of the paper is organized as follows. In Sec.~\ref{general}, we present the
Hamiltonian in a general waveguide-atom coupled system, and give the scattering amplitudes
based on Lippman-Schwinger equation. With these results, we discuss the single-photon scattering and frequency conversion in 1D CRW
with cosine dispersion relation and in optical waveguide with linear dispersion
relation in Sec.~\ref{CRW} and Sec.~\ref{linear},
respectively. In Sec.~\ref{Sec:4}, we draw the conclusions.


\section{\label{general} Formalism of the frequency converter}



\begin{figure}[tbp]
\begin{centering}
\includegraphics[bb=85 270 512 558,width=7cm]{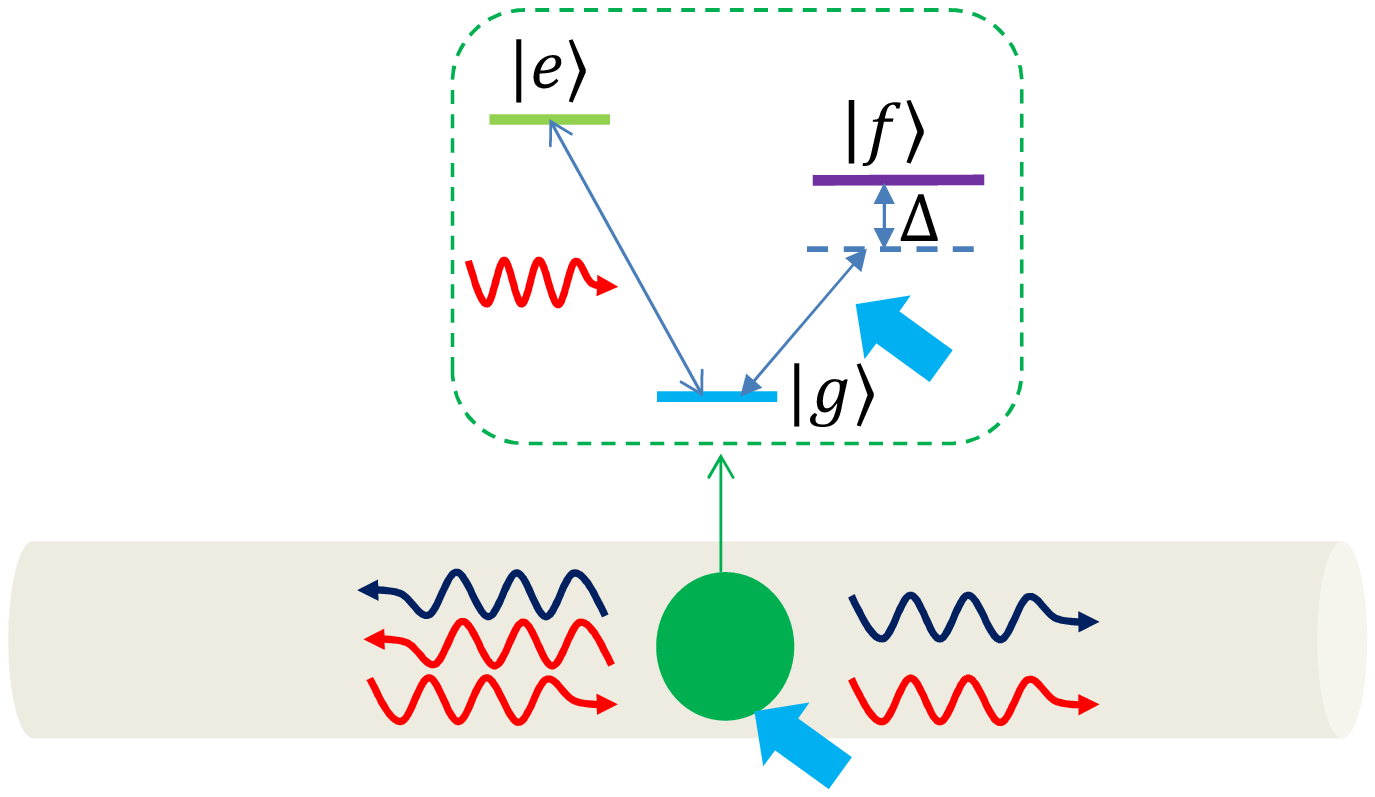}
\end{centering}
\caption{(Color online) The schematic diagram of the model. A $V$-type
atom is located in a 1D waveguide, serving as a frequency converter
for the single photon injected from the left side of the waveguide.}
\label{scheme}
\end{figure}
\subsection{The model and Hamiltonian}


As shown in Fig.~\ref{scheme}, the system we consider contains a 1D waveguide and a $V$-type atom. The $V$-type atom is characterized by a ground state $|g\rangle $, an intermediate state $|f\rangle $ and an excited state $|e\rangle $,
whose energies are denoted as $\omega _{g}$, $\omega _{f}$ and $\omega _{e}$
respectively. The energy of state $|g\rangle $ is set to zero as reference.
The $k$th electromagnetic mode in the waveguide couples to the transition $
|g\rangle \leftrightarrow |e\rangle $ with the strength $J_{k}$, while the
classical driving field with frequency $\nu $ drives the transition $
|g\rangle \leftrightarrow |f\rangle $ with the Rabi frequency $\eta $.
In the rotating frame with respective $H_{0}^{\prime }=v|f\rangle \langle f|$, the Hamiltonian (with $\hbar =1$) of the system is written as
\begin{eqnarray}
H &=&\sum_{k}\omega _{k}a_{k}^{\dagger }a_{k}+\omega _{e}|e\rangle \langle
e|+\eta (|g\rangle \langle f|+|f\rangle \langle g|)  \notag \\
&&+\Delta |f\rangle \langle f|+\sum_{k}(J_{k}a_{k}|e\rangle \langle
g|+J_{k}^{\ast }a_{k}^{\dagger }|g\rangle \langle e|), \label{2-H1}
\end{eqnarray}%
where $a_{k}$ is the annihilation operators for the $k$th mode of
the electromagnetic field with frequency $\omega_k$ in the waveguide.
The dispersion relation between the frequency $\omega_k$ and the wave vector $k$ depends on
the realistic physical system. $\Delta \equiv \omega _{f}-\nu $ is the detuning between the atomic
transition $|g\rangle \leftrightarrow |f\rangle $ and the classical field.

The coupling between  the classical field and the transition $|g\rangle \leftrightarrow |f\rangle $ forms two dressed states
\begin{subequations}
\label{2-H2}
\begin{eqnarray}
|\phi _{+}\rangle  &=&\sin \frac{\theta }{2}|g\rangle +\cos \frac{\theta }{2}|f\rangle ,   \\
|\phi _{-}\rangle  &=&-\cos \frac{\theta }{2}|g\rangle +\sin \frac{\theta }{2}|f\rangle
\end{eqnarray}%
with the corresponding eigenenergies
\end{subequations}
\begin{equation}
\nu _{\pm }=\frac{\Delta \pm \sqrt{\Delta ^{2}+4\eta ^{2}}}{2},  \label{2-H3}
\end{equation}%
where $\tan\theta =2\eta /\Delta $. In the dressed-state representation, the Hamiltonian can be separated as $H=H_{0}+V$: the free Hamiltonian
\begin{equation}
H_{0}=\sum_{k}\omega _{k}a_{k}^{\dagger }a_{k}+\omega _{e}|e\rangle \langle
e|+\sum_{i=\pm }\nu _{i}|\phi _{i}\rangle \langle \phi _{i}|,
\label{2-H4}
\end{equation}%
for the waveguide and the atom, and the atom-waveguide interaction
\begin{equation}
V=\sum_{k}[J_{k}a_{k}(\sin \frac{\theta }{2}|e\rangle \langle \phi
_{+}|-\cos \frac{\theta }{2}|e\rangle \langle \phi _{-}|)+h.c.  \label{2-H5}
\end{equation}%

\subsection{The Lippman-Schwinger equation and scattering amplitudes}


For a photon propagating in the 1D waveguide, when it is far
away from the atom, the system is in free particle states governed by
Hamiltonian $H_{0}$. A photon coming toward the atom will first disappear,
and appear later, i.e., the propagating photon is scattered by the atom.
Taking the state $|k,\phi_n\rangle \equiv a_{k}^{\dagger }|0\rangle\otimes|\phi _{n}\rangle $ ($|0\rangle $ represents the photonic vacuum state in the waveguide and $n=\pm $) as the input state, which represents the atom in the internal state $|\phi _{n}\rangle $ and the photon in the $k$th mode, the
stationary state $|\psi_{kn}\rangle$ can be given by the Lippman-Schwinger equation~\cite{Taylor,JFHuang,DZ2}%
\begin{equation}
|\psi_{kn}\rangle =|k,n\rangle +\frac{1}{E-H_{0}+i0^{+}}V|\psi_{kn}\rangle .
\label{2-S2}
\end{equation}%

Since the excitation number $N=\sum_k a_k^{\dagger}a_k+|e\rangle\langle e|$  is conserved in this system, the eigenstate with one excitation can be written as%
\begin{equation}
|\psi_{kn}\rangle =\sum_{p}( \alpha _{p}|p,\phi _{-}\rangle
+\beta _{p}|p,\phi _{+}\rangle ) +u_{kn}|0,e\rangle .
\label{2-S1}
\end{equation}%
Here, $u_{kn}$ is the probability amplitude of the atom in the
excited state, $\alpha _{p}$ ($\beta _{p}$) is the amplitude for finding one output
photon with wave vector $p$ in the waveguide and the atom in the state $|\phi _{-}\rangle $
$(|\phi_+\rangle)$.

Combining Eqs.~(\ref{2-S1}) and (\ref{2-S2}), and removing the photonic amplitudes,
one can get the amplitude for the atom in the excited state
\begin{equation}
u_{kn}=\frac{\left( \sin \frac{\theta }{2}\delta _{n,+}-\cos \frac{\theta
}{2}\delta _{n,-}\right) J_{k}}{\omega_{kn}-\omega _{e}-\sin ^{2}\frac{\theta }{2}A_{+}-\cos
^{2}\frac{\theta }{2}A_{-}+i0^{+}}
\end{equation}%
with
\begin{equation}
A_{\pm }=\sum_{p}\frac{\left\vert J_{p}\right\vert ^{2}}{\omega_{kn}-\omega _{p\pm
}+i0^{+}},  \label{2-S4}
\end{equation}%
where we have defined $\omega_{kn}:=\omega _{k}+\nu _{n}$.

In the following, we denote the state $|k',\phi_l\rangle=a_{k'}^{\dagger}|0\rangle\otimes|\phi_l\rangle $ $(l=\pm)$ as
the output state, i.e., a photon with wave vector $k'$ is excited in the waveguide and
the atom is in the internal state $|\phi_l\rangle $. Since the atom is in one of the dressed states
in the presence of a photon of frequency $\omega _{k}$ before the scattering process, during the process it will absorb the photon with
frequency $\omega _{k}$ and pass into the excited state $\left\vert e\right\rangle $, then emit the photon
with frequency $\omega _{k^{\prime }}$ and pass into any of the dressed states. The conservation of the total energy in the incoming and
outgoing states offers a possibility for the outgoing photon carrying a
frequency different from the absorbed one. After the scattering process, if the
atom is found in the same state as the input state, i.e., $|\phi _{l}\rangle=|\phi _{n}\rangle $,
we will have $\omega _{k^{\prime }}=\omega _{k}$. However, if $|\phi _l\rangle\neq |\phi _n\rangle $ the frequency of
photon will be changed ($\omega _{k^{\prime }}\neq \omega _{k}$) since the eigenenergies of $|\phi _{\pm }\rangle$ are different. The evidence on the conservation of energy in the scattering process can be found in the
following element of the $S$-matrix%
\begin{equation}
S_{k',l\leftarrow k,n}=
\delta _{l,n}\delta _{k,k^{\prime }}-2\pi i\delta (\omega _{k^{\prime
},l}-\omega _{k,n})T_{k';l\leftarrow k;n},
\label{2-S5}
\end{equation}
where the elements of the on-shell $T$ matrix are obtained as~\cite{Taylor}
\begin{eqnarray}
T_{k',l\leftarrow k,n}&\equiv&\langle k',\phi_l|V|\psi_{kn}\rangle\\ \nonumber&=&u_{kn}J_{k^{\prime }}^{\ast }(\sin \frac{\theta }{2}\delta
_{l,+}-\cos \frac{\theta }{2}\delta _{l,-}).
\end{eqnarray}
In the above, the absorption and emission of single photons by the atom
is formalized as a multi-channel scattering process. Here,
there are two open channels with available states $|k,\phi _{-}\rangle$ and $
|k,\phi _{+}\rangle$, respectively. Hereafter, we
will call them ``negative channel" and ``positive channel"  according to the
related atomic states $\left\vert \phi _{-}\right\rangle$ and $\left\vert \phi _{+}\right\rangle $, respectively.
In the following discussion, we restrict our consideration to
the case that the single photon is incident from the negative channel with
wave vector $k$($>0$), then the element of the $S$-matrix in the negative channel is
\begin{equation}
S_{k',-\leftarrow k,-}=r_{-}\delta _{k^{\prime },-k}+t_{-}\delta _{k^{\prime
},k}  \label{2-S6}
\end{equation}%
where $r_{-}$ ($t_{-}$) is the reflection (transmission)
amplitude, and $t_{-}=r_{-}+1$. The element of the $S$-matrix in the positive channel is
\begin{equation}
S_{k',+\leftarrow k,-}=t_{+}[\delta _{k',q(k)}+\delta _{k',-q(k)}],
\label{2-S7}
\end{equation}%
where the forward and backward transfer amplitudes are equal and denoted by $t_{+}$.
For the sake of simplicity, in what follows we will write the wave vector $q(k)$($>0$) as $q$,
whose dependence on the input wave vector $k$ is given by the implicit relation
\begin{equation}
\omega_{q}=\omega_{k}-\nu_++\nu_-.
\label{conser}
\end{equation}
It shows from Eqs.~(\ref{2-S6},\ref{2-S7}) that, the wave vector of the
scattering photon $k'$ satisfies $k'=\pm k$ when the incident photon is confined in the negative
channel and $k'=\pm q$ when it is transferred to the positive channel.

It can be found from Eq.~(\ref{2-S5}) that when the photonic flow is confined to the incident channel,
the frequency of the emitted photon is equal to the absorbed one. However,
when the incident photon is transferred to another channel, the frequency of the
emitted photon will be lowered or raised by the amount $\left\vert \nu
_{+}-\nu _{-}\right\vert $. Consequently, the atom acts as a frequency
converter for single photons propagating in the 1D waveguide.


\section{\label{CRW} Single-photon scattering in CRW}


A 1D CRW is typically made of single-mode resonators that are coupled to
each other through the evanescent tails of adjacent mode function of the
cavity field. In experiments, the 1D CRW can be realized in photonic
crystal, where the atom can be realized by a defect~\cite{KHen}. With the
improvement of the fabrication techniques, a much stronger coupling between
the atom and the waveguide mode has been realized in the superconductor
transmission line, where the natural atom is also replaced by the
superconductive qubit~\cite{JMa}. Assuming that all the resonators have the same
frequency $\omega $ and the intercavity couplings $\xi $ between any two
nearest-neighbor cavities are the same, the 1D CRW is characterized by the
dispersion relation%
\begin{equation}
\omega _{k}=\omega -2\xi \cos kl
\label{3-CRW1}
\end{equation}%
where $l$ is the lattice constance. In the rest of this section, the wave vector $k$ is
dimensionless by setting $l=1$. For an atom located in the $a$th resonator, the atom-cavity coupling
strength is $J_{k}=Je^{ika}$. Here, $J$ is assumed to be real.
\begin{figure}
\begin{centering}
\includegraphics[bb=67 331 560 650,width=8cm]{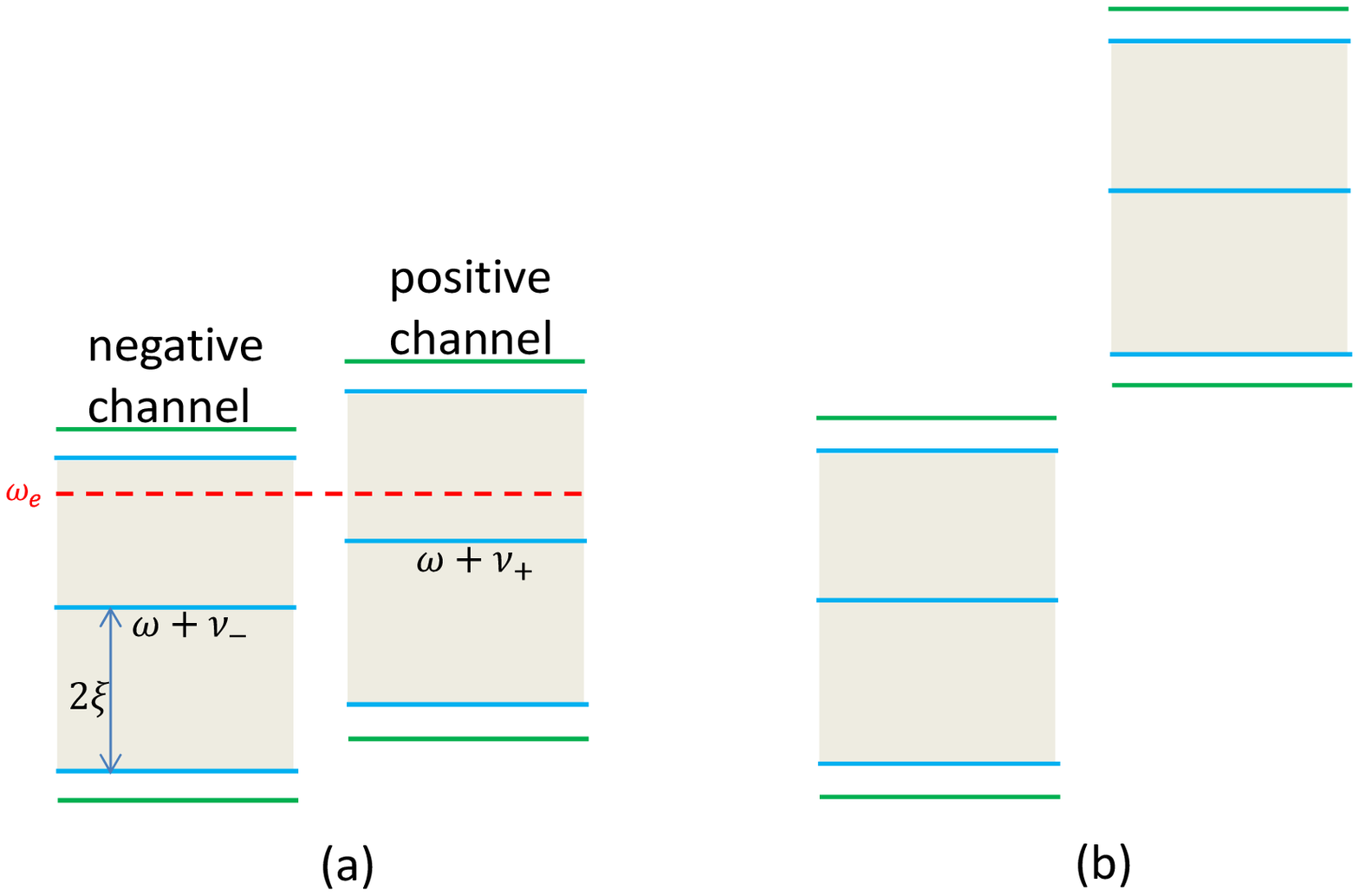}
\end{centering}
\caption{(Color online) The energy band configurations for the two channels
in the CRW. (a) The energy bands are partially overlapped. (b) The energy
bands are completely separated. The thick lines represent the bound states in the
corresponding channels and the red dashed line represents the energy of the
atomic excited state $|e\rangle $.}
\label{cosin}
\end{figure}

The energies for the free particle states in the negative (positive) channel form an energy
band with the bandwidth $4\xi $, which is centralized at $\omega +\nu _{-}$ ($\omega +\nu _{+}$).
The broken translation symmetry of the CRW allows
bound states below and above the energy band of each channel~\cite{Lan13}.
Since $\nu _{-}\neq \nu _{+}$, by adjusting the strength and the frequency of
the driving field, we could achieve two following band configurations as shown
in Fig.~\ref{cosin}: (a) partial overlap between the two bands; (b) no overlap
between them.

\begin{figure}[tbp]
\begin{centering}
\includegraphics[width=8cm]{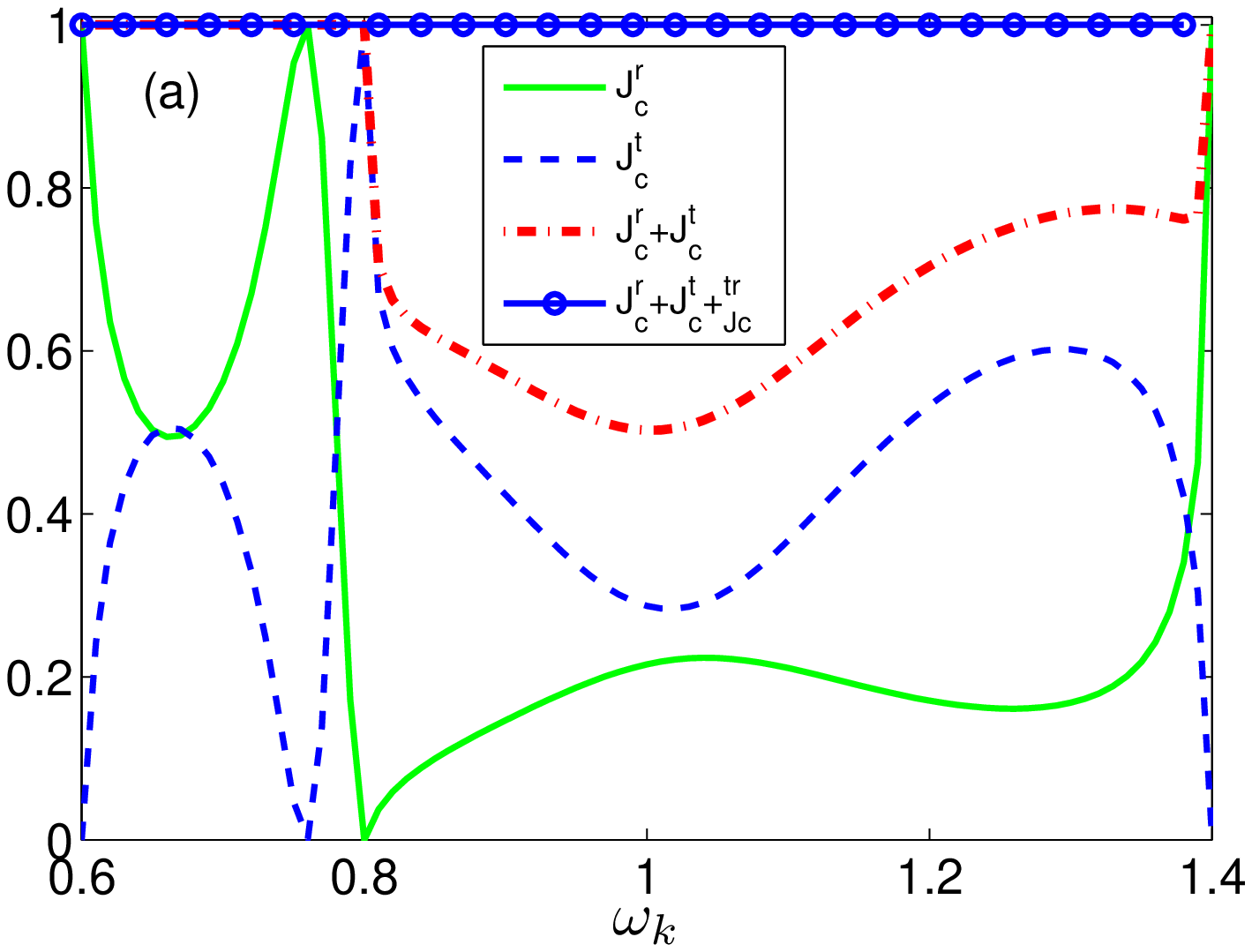}
\par\end{centering}
\begin{centering}
\includegraphics[width=8cm]{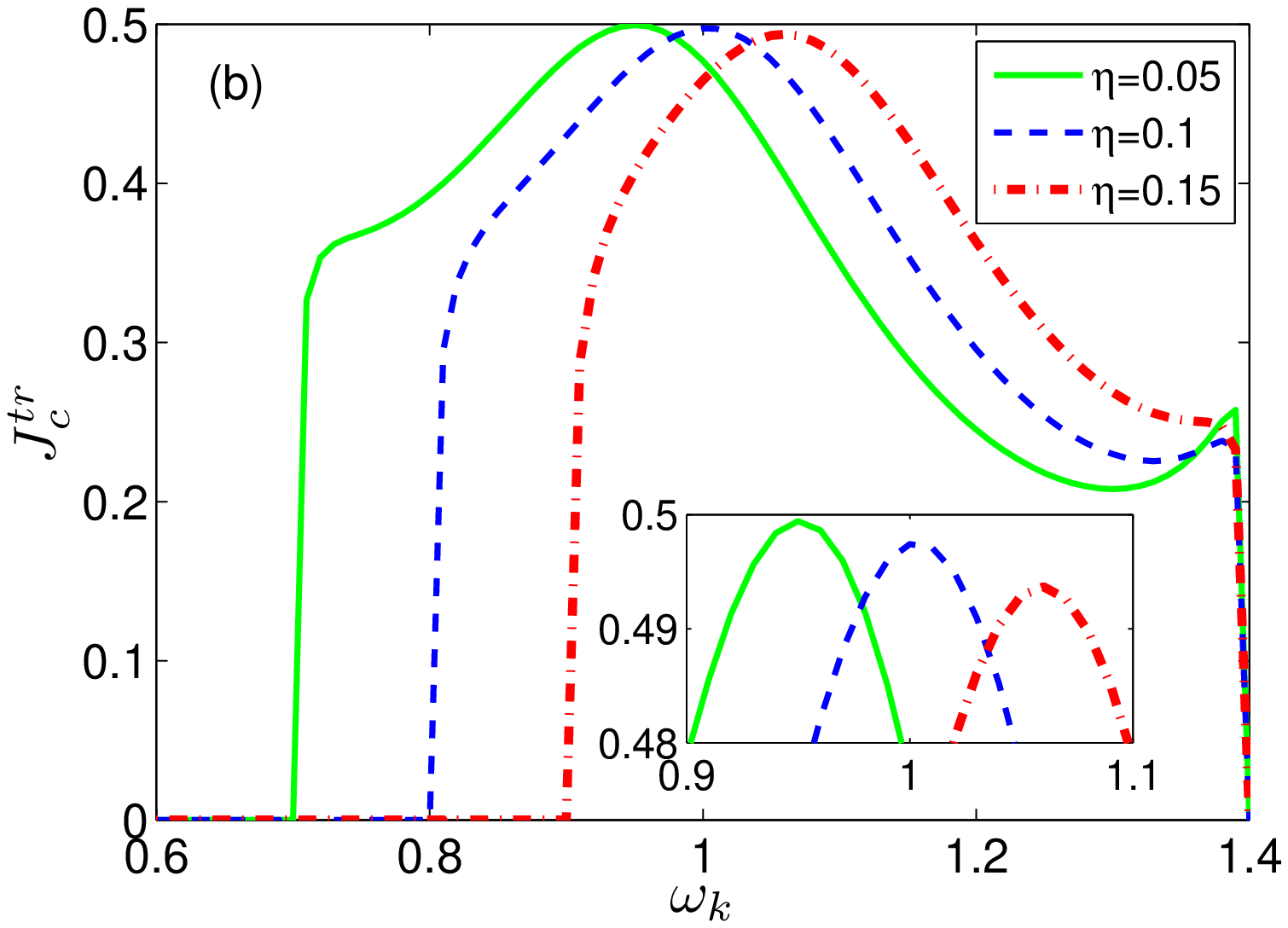}
\par\end{centering}
\begin{centering}
\includegraphics[width=8cm]{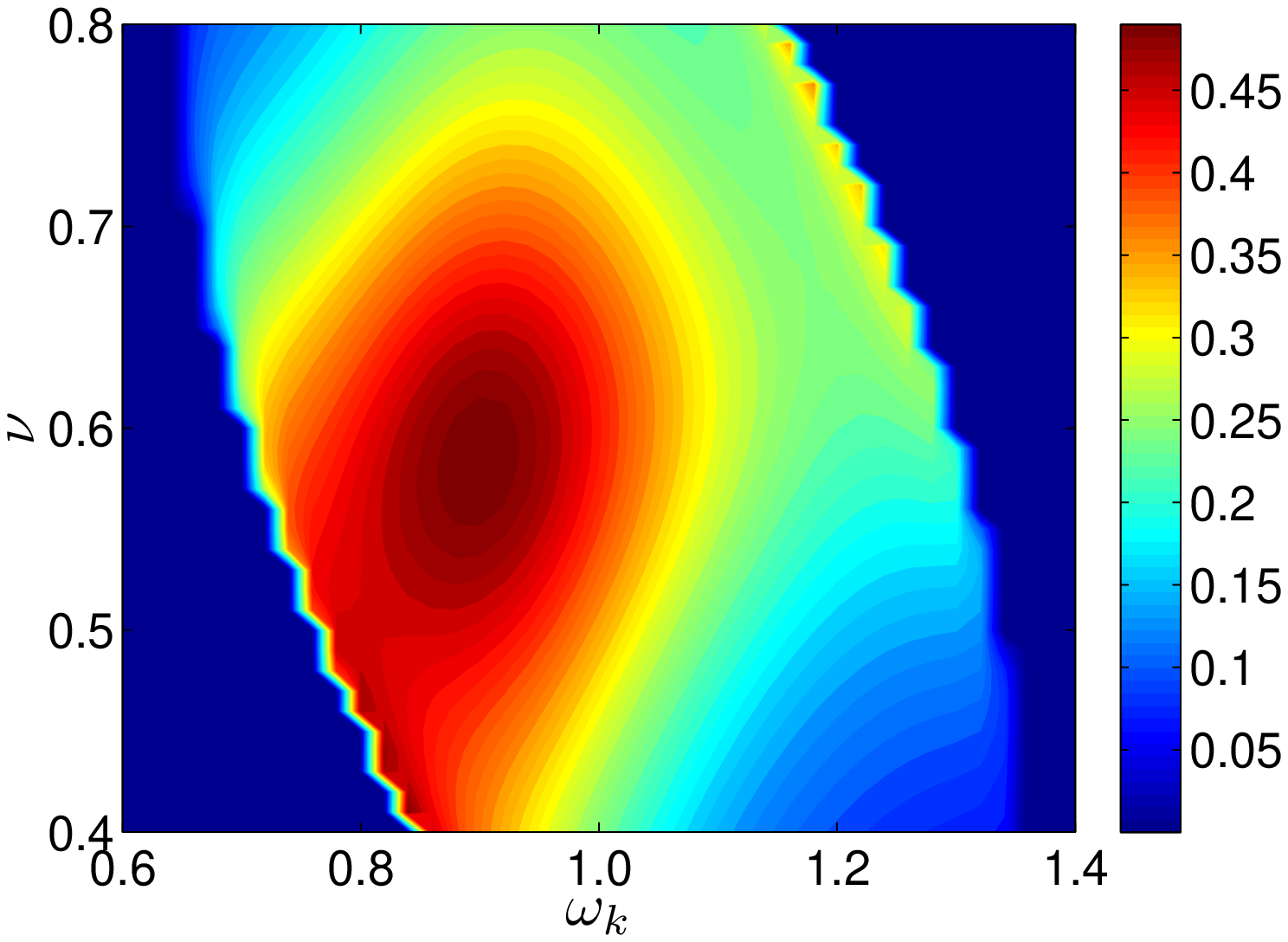}
\par\end{centering}
\caption{(Color online) The scattering flows (in units of the incident flow) as a function of the incident
frequency in CRW when the two energy bands are partially overlapped.
(a) The reflection and transmission flows when $\Delta=0$ and $\eta=0.1$.
(b) The transfer flows for different $\eta$ under the resonance situation $\Delta=0$. (c) The
contour of the transfer flow versus $\nu$ and $\omega_{k}$ with $\omega_f=0.6$
and $\eta=0.1$. The other parameters are set as $\xi=0.2$, $\omega_e=0.9$ and $J=0.3$ in units of $\omega=1$.}
\label{cosinflow}
\end{figure}

For a propagating state excited in the negative channel, the reflection
and transfer amplitudes are obtained from Eqs.~(\ref{2-S6},\ref{2-S7}) utilizing
the residue theorem as
\begin{subequations}
\label{3-CRW2}
\begin{eqnarray}
r_{-}^{c}&=&\frac{\cos^{2}\frac{\theta}{2}}{2i\xi(\frac{\omega_{k-}-\omega_{e}}{J^2}+\frac{i\sin^{2}\frac{\theta}{2}}{2\xi\sin q})\sin k-\cos^{2}\frac{\theta}{2}}\notag \\ \\
t_{+}^{c}&=&\frac{e^{i(|q|-k)a}\cos\frac{\theta}{2}\sin\frac{\theta}{2}}{2i\xi(\frac{\omega_{k-}-\omega_{e}}{J^2}
+\frac{i\cos^{2}\frac{\theta}{2}}{2\xi\sin k})\sin q-\sin^{2}\frac{\theta}{2}}\notag \\
\end{eqnarray}
\end{subequations}
where the superscript ``$c$" refers to the quantum channel made of ``CRW"
and the subscript $-$ ($+$) refers to the negative (positive) channel.

The cosine type dispersion in Eq.~(\ref{3-CRW1}) characterizes the nonlinear
relation between the frequency and the wave vector of the traveling photon in CRW.
Therefore, for the incident photon with a fixed wave vector $k$, the
group velocity is  $v_g=2\xi \sin{k}$. Meanwhile, the scattered photon
in different channels will posse different group velocities. That is,
$v_g=2\xi \sin{k}$ in the negative channel and $v_g=2\xi \sin{q}$ in the positive channel,
where the wave vector $q$ is defined in Eq.~(\ref{conser}). Attentions should be paid here
that the group velocities have the same unit with frequency.

In this sense, we define the scattering flows of the single photons as the square modulus of the scattering amplitudes
multiplying the group velocities in the corresponding channels. For the sake of simplicity, we
set the incident flow as unit and the reflection (transmission) flow in the
negative channel is calculated as $J_{c}^{r}(J_{c}^{t})=|r_{-}^{c}(t_{-}^{c})|^{2}$
and the transfer flow to the positive channel is obtained as $J_{c}^{tr}=2|t_{+}^{c}|^{2}\sin q/\sin
k$ or $J_{c}^{tr}=0$ depending on whether the energy of the incident photon
is inside or outside the energy band of the positive channel. It then follows
from Eqs.~(\ref{3-CRW2}) that the scattering flows are independent of the atom position.

Consider the case where the two energy bands are partially overlapped. Here,
the incident frequency may be either inside or outside the continuous regime
of the positive channel. In Fig.~\ref{cosinflow}(a), we plot the reflection
flow $J_{c}^{r}$, the transmission flow $J_{c}^{t}$, the summation $J_{c}^{r}+J_{c}^{t}$
in the negative channel, and the total flow $J_{c}^{r}+J_{c}^{t}+J_{c}^{tr}$
as a function of the incident frequency. It can be observed that the single photon
is confined to the negative channel when the energy of the incident state is
out of the overlap region of the two continuum bands. Consequently, the flow's
conservation is described by $J_{c}^{r}+J_{c}^{t}=1$. When the incident energy
is within the overlap region of the two continuum bands, the flow in the negative
channel satisfies $J_{c}^{r}+J_{c}^{t}<1$, which means the incident photons can be transferred
to the positive channel, and the flow conservation is changed as $J_{c}^{r}+J_{c}^{t}+J_{c}^{tr}=1$.
One can also find the complete reflection when the single photons are confined
to the negative channel, which is the result of the interference between the open
channel provided by the negative channel and the close channel provided by the
positive channel. This close channel is formed by a bound state, which is the
consequence of the atom breaking the translation symmetry of the positive channel.
The energy of the bound state in the positive channel can be obtained by the
transcendental equation
\begin{equation}
\omega_{k-}-\omega_{e}=\frac{J^{2}\sin^2(\theta/2)}{2i\xi\sin q}.
\end{equation}
Actually, the bound state energy of the positive channel corresponds to the poles
of the scattering matrix of the system studied in Ref.~\cite{Lan08}, i.e., a
two-level system embedded in the positive channel.

In Fig.~\ref{cosinflow}(b), we plot the transfer flow $J_{c}^{tr}$ as
a function of the frequency of the incident photon in the negative channel
when the driven field is resonant
with the atomic transition $|g\rangle \leftrightarrow |f\rangle$. It can
be observed that when the frequency of the incident photon is resonant with
the transition frequency between the atomic excited state $|e\rangle$ and
the dressed state $|\phi_{-}\rangle$ ($\omega_{k}=\omega_e-\nu_-$), the
transfer flow reaches its maximum. As the Rabi frequency $\eta$ increases,
the maximum of the transfer flow decreases monotonically. When the driving
field is strong enough so that the two energy bands are completely separated
(as shown in Fig.~\ref{cosin}(b)), the transfer flow vanishes, i.e., the
single photons can not travel in the positive channel. In Fig.~\ref{cosinflow}(c),
we have plotted the transfer flow versus the frequency $\nu$ of the driven field
and the incident frequency $\omega_{k}$ by fixing the transition frequency
$\omega_f=0.6$ and the Rabi frequency $\eta=0.1$ in units of $\omega =1$. It can be found that the
transfer flow reaches its maximum  only when the frequency of the classical
field is resonant with the atomic transition $|g\rangle\leftrightarrow |f\rangle$.

The above discussion shows that the overlap of the two bands is necessary for
the atom to fulfill the function of a photonic frequency converter. Here, a
weak driving field is preferred so that the frequency difference in the two
bands can not be too large. However, the preference for weak driving field is
not necessary when the CRW is replaced by a 1D waveguide with a linear dispersion
relation, which will be discussed in the next section.


\section{\label{linear} Single-photon scattering in a one-dimension optical
waveguide}


In this section, we focus on the single-photon scattering in a 1D
optical waveguide which naturally arises in a nanophotonic system. It exhibits
a linear dispersion relation $\omega_{k}=v_{g}|k|$, where $k$ is the wave
vector and $v_{g}$ is the group velocity of the light in the waveguide. We consider that the $V$-type
atom is located at the position $x=a$ and the coupling strength between the
atom and the waveguide is $J_{k}=Je^{ika}$.

\begin{figure}[tbp]
\begin{centering}
\includegraphics[bb=80 530 424 770,width=7cm]{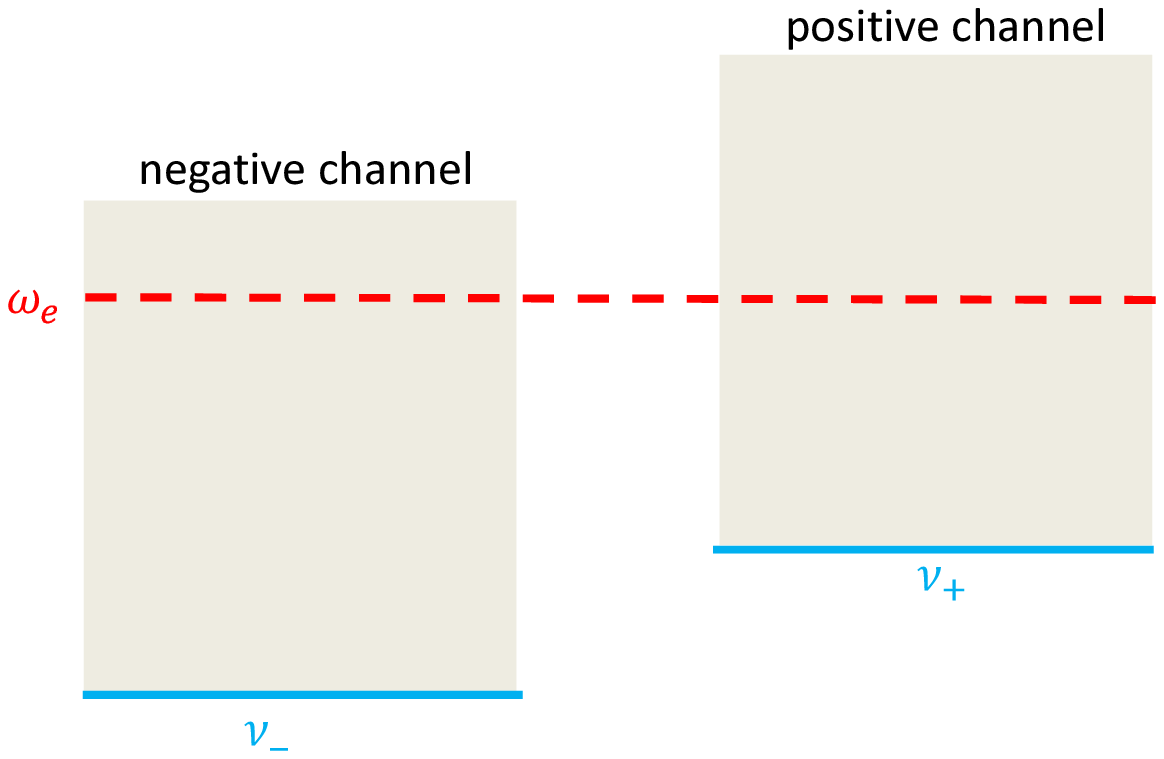}
\end{centering}
\caption{(Color online) The energy level configurations for the two channels
in the optical waveguide. The red dashed line represents the frequency of the
atomic excited state. }
\label{lin}
\end{figure}

The energy of the free particle states in the negative (positive) channel have the lower
energy bounds at $\nu_-$ and $\nu_+$ respectively, but without the upper
bounds, as shown in Fig.~\ref{lin}. For single photons with the wave vector $k$
incident from the negative channel, the reflection and transfer amplitudes
in the optical waveguide are calculated from Eqs.~(\ref{2-S6},\ref{2-S7})
utilizing the residue theorem as
\begin{subequations}
\begin{eqnarray}
r_{-}^{o}&=&\frac{J^{2}\cos^{2}\frac{\theta}{2}}{iv_{g}(\omega_{k-}-\omega_{e})/L-J^{2}},\\
t_{+}^{o}&=&\frac{J^{2}e^{i(|q|-k)a}\cos\frac{\theta}{2}\sin\frac{\theta}{2}}{iv_{g}(\omega_{k-}-\omega_{e})/L-J^{2}}\label{tro}.
\end{eqnarray}
\end{subequations}
Here, the superscript ``$o$" implies the quantum channel made of ``optical" waveguide
and $L$ is the length of the waveguide. The relationship between the wave vector $q$ in
positive channel and the wave vector $k$ in the negative channel is obtained from Eq.~(\ref{conser}) as
\begin{equation}
|q|=|k|-\frac{\nu_+-\nu_-}{v_g}.
\end{equation}

Similar to the last section, we can also define the scattering flows. Since the
group velocities in both of the channels are the same, the reflection (transmission) flow
in the negative channel is $J_{o}^{r} (J_{o}^{t}) =|r_{-}^{o}(t_{-}^{o})|^{2}$, and
the transfer flow in the positive channel is $J_{o}^{tr}=2|t_{+}^{o}|^{2}$. It is same
with the case in CRW that the scattering flows are independent of the atom position.

\begin{figure}[tbp]
\begin{centering}
\includegraphics[width=8cm]{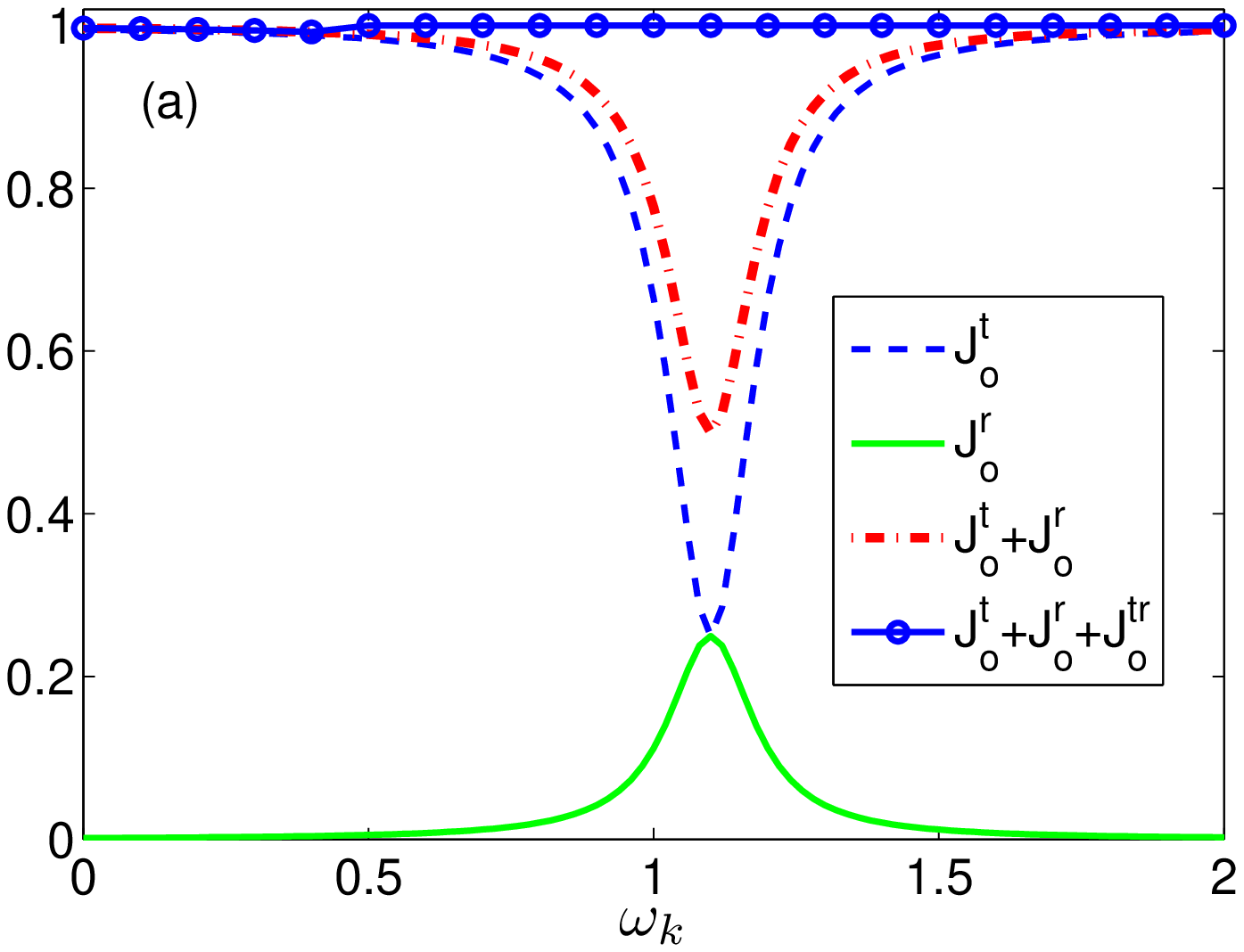}
\par\end{centering}
\begin{centering}
\includegraphics[width=8cm]{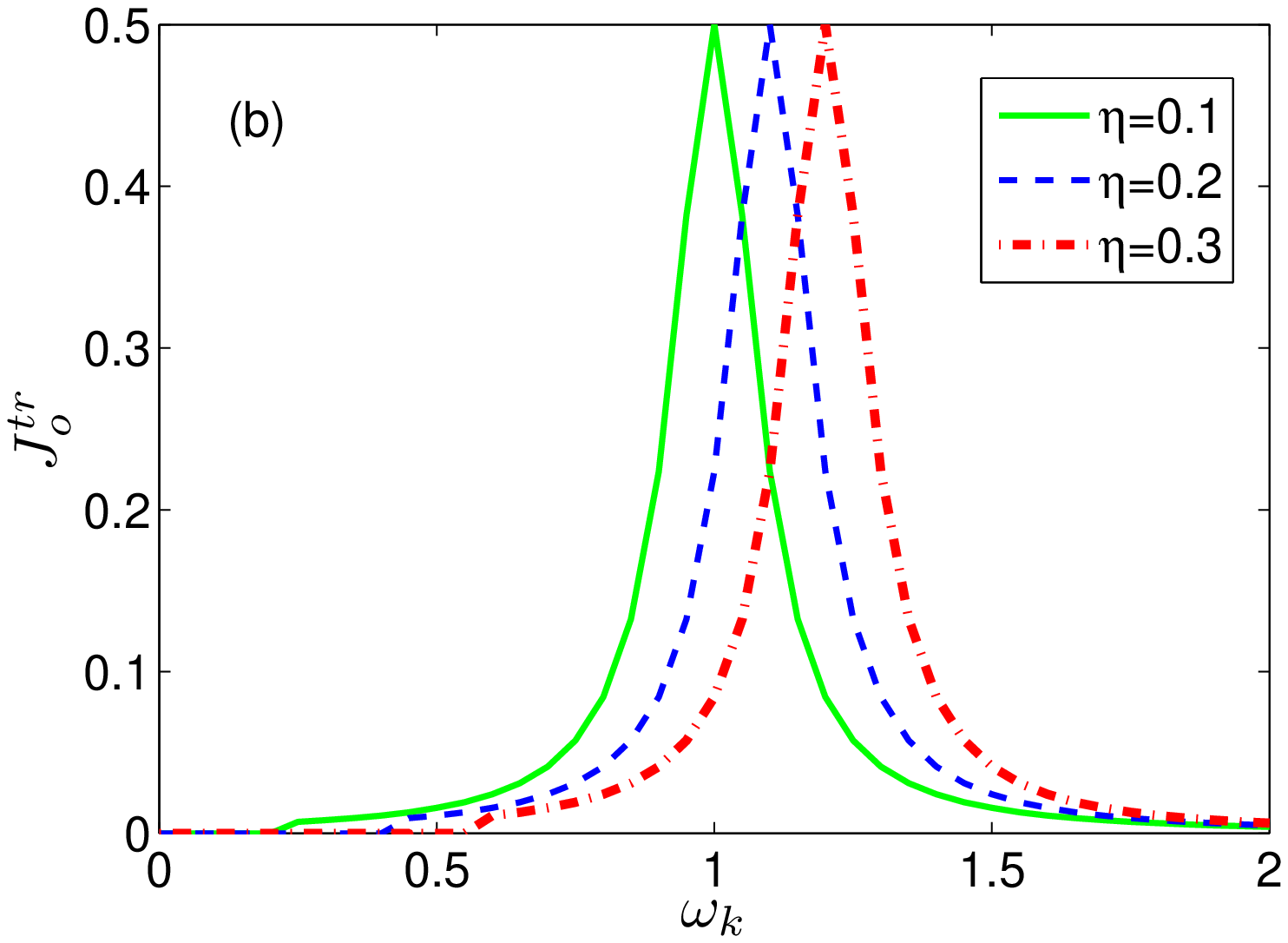}
\par\end{centering}
\caption{(Color online) The scattering flows (in units of the incident flow) as a function of the incident
frequency in the optical waveguide. The parameters are set as $\omega_e=0.9$, $J=0.3$, and $\Delta=0$.
All of the parameters are in units of $v_g/L=1$. Under these parameters,
the energy for the atom dressed ground states are $\nu_{\pm}=\pm\eta$. (a) The reflection and transmission flows when $\eta=0.2$. (b) The transfer flows for different $\eta$.}
\label{oflow}
\end{figure}

In Fig.~\ref{oflow}, we plot the scattering flows as a function of the incident frequency
with the assumption that the classical field resonantly drives the atomic transition. The
reflection flow, transmission flow, the flow in the negative channel, and the total flow in the
system is displayed in Fig.~\ref{oflow}(a)and the transfer flow is depicted in Fig.~\ref{oflow}(b).
It can be observed from Fig.~\ref{oflow}(a) [also in Fig.~\ref{oflow}(b)] that when the incident
frequency is below the difference between the lower
bound of the positive and negative channels ($\omega_k<\nu_+-\nu_-=2\eta$), the photon is confined in the negative channel. Consequently,
$J_{o}^{tr}=0$ and the flow conservation equation is $J_{o}^{r}+J_{o}^{t}=1$. However, the single
photon is transferred to the positive channel once the incident frequency surpasses the energy difference
 of the lower bound of the two channels and the flow conservation equation becomes
$J_{o}^{r}+J_{o}^{t}+J_{o}^{tr}=1$. It is observed from Fig.~\ref{oflow}(b) that, the transfer
flow achieves its maximum with magnitude $0.5$ when the incident photon is resonant with the
$|e\rangle \leftrightarrow |\phi_{-}\rangle$ transition, i.e., $\omega_k=\omega_{e}-\nu_-$.
The maximum value of the transfer flow is independent of the strength of the driving light.
When the frequency of the incident photon is far off resonance from the atomic transition, the
transfer flow is close to zero, and the photon is nearly perfectly transmitted [as shown in
Fig.~\ref{oflow}(a)]. When the classical field is off-resonant with the atomic transition
$|g\rangle \leftrightarrow |f\rangle$, the maximum of the transfer flow is lowered, i.e.,
always smaller than $0.5$, which can be understood from Eq.~(\ref{tro}) and the relation
$\tan\theta=2\eta/\Delta$. For a resonant classical field, the detuning vanishes, and $\theta=\pi/2$,
then the value of the numerator in Eq.~(\ref{tro}) achieves its maximum. For a off-resonant
classical field $\nu\neq\omega_f$, $\theta<\pi/2$. Consequently, the value of the numerator
in Eq.~(\ref{tro}) is lowered.

The above discussion shows that the atom functions as a photonic frequency converter in 1D optical
waveguide as long as the energy of the incident state lies inside the overlap region of the two channels.
For single photons incident from the negative channel, it is possible to find that the carried frequency
of the outgoing photon is red shifted. However, single photons incident from the positive channel can be
also transferred to the negative channel,with the carried frequency of the outgoing photon being blue shifted instead of red shifted. We note that there is no total reflection below the lower bound of the positive channel, which means
that there is no bound state for single photons in a waveguide with the linear dispersion relation.
The requirement for converting the single-photon frequency is that the lower bound of the positive
channel is below the transition frequency $\omega_e$.

\section{\label{Sec:4}Conclusion}

In this paper, we have studied the absorption and re-emission process of single photons by a $V$ type atom
embedded in a 1D waveguide. To show how the $V$-type atom behaves as a single-photon frequency converter
by applying a classical field to drive the transition between the ground and intermediate state, we
formulate the absorption and re-emission process as a two-channel scattering process. We analytically
investigate the scattering flows in a 1D CRW with cosine dispersion and an optical waveguide with linear
dispersion. It is found that converting the frequency of single photons in a 1D waveguide requires
that (1) there is overlap between the continuums of the two channels; (2)the energy of the atom excited state is
in the continuums of the two channels. As long as the requirements are satisfied, the strength of the classical
field can be arbitrary large for the waveguide with linear dispersion relation to achieve the maximal
probability of the outgoing photon with different frequency, however, a weak classical field is
preferred in the 1D CRW.

\begin{acknowledgments}
We thank the helpful discussions from D. Z. Xu, L. Ge, S. W. Li and
X. W. Xu. This work is supported by the NSFC
(under Grant No. 11174027) and the National 973 program
(under Grant No. 2012CB922104 and No. 2014CB921402). L. Zhou is supported by NSFC No.~11374095,
NFRPC~2012CB922103 and Hunan Provincial NSFC 12JJ1002.

\end{acknowledgments}

\end{document}